%% file: main.tex
\documentclass{llncs}
\usepackage{amsmath}
\usepackage{todonotes}
\usepackage{lscape}
\usepackage{amssymb}
\usepackage{amsfonts}
\usepackage{stmaryrd}
\usepackage{array}
\usepackage{subfig}
\usepackage{graphicx}
\usepackage[ruled]{algorithm2e}
\usepackage[inference]{semantic}
\input{macros.tex}
\usepackage{tikz}
\input{figures/tikz/tikz.tex}
\usepackage{mathtools}
\usepackage{csquotes}
\begin{document}
\title{Aggressive Aggregation: a New Paradigm \\
for Program Optimization}
\author{Frederik Gossen \and
  			Marc Jasper \and	
  			Alnis Murtovi  \and
  			Bernhard Steffen}
\institute{TU Dortmund University, Germany \\
 \email{\{firstname.lastname\}@tu-dortmund.de}}
\maketitle
\begin{abstract}
  In this paper, we propose a new paradigm for program optimization which is based on aggressive aggregation, i.e., on a partial evaluation-based decomposition of acyclic program fragments into a pair of computationally optimal structures:
  an Algebraic Decision Diagram (ADD) to capture conditional branching and a parallel assignment that refers to an Expression DAG (ED) which realizes redundancy-free computation. The point of this decomposition into, in fact, side-effect-free component structures allows for powerful optimization that semantically comprise effects traditionally aimed at by SSA form transformation, code specialization, common subexpression elimination, and (partial) redundancy elimination.
  We illustrate our approach along an optimization of the well-known iterative Fibonacci program, which, typically, is 
  considered to lack any optimization potential. 
  The point here is that our technique supports loop unrolling as a first class optimization technique and is tailored to optimally aggregate large program fragments, especially those resulting from multiple loop unrollings. 
  For the Fibonacci program, this results in a performance improvement beyond an order of magnitude.
  
\end{abstract}

\section{Introduction}
\label{sec:intro}
Traditionally, program optimization techniques like redundancy elimination~\cite{morelRenvoise,poe,partialDeadCodeElimination}, code motion~\cite{optimalCodeMotion}, strength reduction~\cite{lazySR}, constant propagation/fold\-ing~\cite{finiteConstants}, and loop invariant code motion~\cite{aho1986compilers} are quite syntax-oriented and mostly preserve program structure. 
Programs such as an iterative implementation of computing the $n$-th Fibonacci number typically remain untouched by such optimizations.

In this paper, we propose a new paradigm for program optimization which is based on aggressive aggregation, i.e. on a partial evaluation-based decomposition of acyclic program fragments into a pair of computationally optimal structures: 
an Algebraic Decision Diagram (ADD) to capture conditional branching and a parallel assignment that refers to an Expression DAG (ED) which realizes redundancy-free computation.

The point of this decomposition is to obtain large program fragments which can be optimized using ADD technology, SMT solving, and expression normalization without being `disturbed’ by side effects. 
Not only are multiple occurrences of a term guaranteed to be semantically equivalent as in SSA form~\cite{ssa},  
but the large size of the arising aggregated expressions further increases the optimization potential.

Our approach enables provably optimal transformations via heuristics in a transparent fashion:
\begin{enumerate}
	\item Cut points, similar to the ones in Floyd’s inductive assertion method~\cite{floyd}, are used to split the program into acyclic fragments (Sec.~\ref{subsec:cutpoints}). 
  This heuristic step may be enhanced by, e.g., loop unrolling to increase the optimization potential (Sec.~\ref{subsec:loop_unrolling}).
	
	\item The resulting acyclic fragments are each decomposed into a \emph{path condition} and a computational part in terms of a \emph{parallel assignment}. 
  This decomposition is executed in a canonical fashion reminiscent of symbolic execution (Sec.~\ref{subsec:decomposition}).
	
	\item The path condition is then transformed into an Algebraic Decision Diagram (ADD) which is symbolically canonical for a fixed predicate order (Sec.~\ref{subsec:path_aggr}). 
  Each terminal of an ADD points to a corresponding pair of parallel assignment and successor cut point.
	
	\item Expressions that occur within the input program and during its decomposition are stored in an Expression DAG (ED) (Sec.~\ref{subsec:expr_dag}).
  Predicates inside the ADD and right-hand sides of parallel assignments both simply reference nodes in the ED.
	
	\item Thereafter, further optimizations are applied such as a removal of infeasible paths in the ADD (Sec.~\ref{subsec:unsat_path_removal}) and a normalization of the ED (Sec.~\ref{subsec:normalization}), both using an SMT solver.
	
\end{enumerate}

Steps $1$ and $5$ are the ones that most strongly depend on good heuristics: 
The choice of good cut points is essential, as are decisions regarding loop unrolling. 
Similarly, expression normalization is a heuristic 
(i)~as the equivalence of arithmetic expressions is in general undecidable and
(ii)~because it has to be adapted to a given setting, e.g. the precise hardware architecture.\footnote{Note that the applicability of normalization rules further depends on the desired semantics, e.g. in case that finite precision arithmetic should be modeled.}
Moreover, infeasible path detection is in general undecidable and depends on the heuristics in the applied SMT solver, in our case Z3~\cite{z3}.
In contrast to that, steps $2$ -- $4$ are mostly canonical. 
Only the ADD construction depends on a good predicate order (Sec.~\ref{subsec:loop_unrolling}).

We will illustrate our approach along an optimization of the well-known iterative Fibonacci program, which, typically, is considered to lack any optimization potential (Sec.~\ref{sec:eval}). 
The point here is that our technique supports loop unrolling as a first class optimization technique: 
It is tailored to optimally aggregate large program fragments---especially those resulting from multiple loop unrollings---while only incurring the expected linear increase in size.
In fact, we are able to achieve a performance improvement during the computation of the $n$-th Fibonacci number of more than an order of magnitude.

The correctness of our program decomposition into side-effect-free fragments (step 2, see also Theorems~\ref{thoerem:correctness_decomposition} and~\ref{thoerem:correctness_segmentation} as well as Corollary~\ref{corollary:correctness_decomposition}) can be elegantly proven via a path-oriented `alignment’ of simultaneous concrete and symbolic execution using the pattern of Structural Operational Semantics (SOS)~\cite{PLOTKIN}, a method that we use as a more general proof principle.

Section~\ref{sec:while_language} briefly introduces the language of programs that we address in this paper.
Thereafter, Section~\ref{sec:decomp} presents our new compilation paradigm based on aggressive aggregation.
Additional optimizations are described in Section~\ref{sec:opt} before Section~\ref{sec:eval} evaluates our approach based on a speedup of the transformed Fibonacci program.
Section~\ref{sec:concl} presents our conclusion and a brief outlook to future work.

\section{Program Language}
\label{sec:while_language}
We choose a simple while language~\cite{while} as a representative formalism for programs that we compile:
\begin{definition}[While Language]
	Let $\var$ be a set of integer variables with $x \in \var$. Let $\ArEx$ be arithmetic expressions and $\BoEx$ Boolean expressions.
	Our while language comprises programs $S$ according to the following Backus-Naur form  (BNF):%
	
	\normalfont
	\begin{align*}
	S \bnfdef &x \assign \ArEx \bnfor \keyword{skip} \bnfor S; ~ S \bnfor \\
	&\keyword{if} ~ \BoEx ~ \{ S \} ~ \keyword{else} ~ \{ S \} \bnfor \keyword{while} ~ \BoEx ~ \{ S \}
	\end{align*}
	\label{def:while_language}
\end{definition}
The arithmetic and Boolean expressions that our while language is based on are defined as follows:

\begin{definition}[Program Expressions]
	\label{def:expr}
	Let $\mathbb{Z}$ denote the domain of integers and $\var$ be a set of integer variables. Arithmetic expressions $\ArEx$, Boolean expressions $\BoEx$, and atomic propositions $\AP$ are defined by the following BNFs:
	\begin{align*}
	\ArEx &\bnfdef \var \bnfor \mathbb{Z} \bnfor \ArEx + \ArEx \bnfor \ArEx - \ArEx \bnfor \ArEx * \ArEx \bnfor \ArEx / \ArEx   \\
	\BoEx &\bnfdef \BoEx \vee \BoEx \bnfor \BoEx \wedge \BoEx \bnfor \neg \BoEx \bnfor \AP                 \\
	\AP &\bnfdef \ArEx < \ArEx \bnfor \ArEx = \ArEx \bnfor true \bnfor false
	\end{align*}	
\end{definition}
We parenthesize expressions implicitly along these BNF rules. 

\begin{algorithm}[!htb]
	\SetKwIF{If}{ElseIf}{Else}{if}{\normalfont \{}{elif}{{\normalfont \}} else {\normalfont \{}}{\normalfont \}}%
	\SetKwFor{While}{while}{\normalfont \{}{\normalfont \}}%
	\SetAlgoNoLine
	\KwIn{Positive integer $n$}
	\KwOut{The $n$-th Fibonacci number $fib$ }
	
	\eIf{$\neg(1 < n)$}{
		$fib \assign 1$ \; 
	}{ % else
		$prev \assign 1$ \; 
		$fib \assign 1$ \;  
		\While{$2 < n$}{
			$tmp \assign prev + fib $ \;
			$prev \assign fib$ \;
			$fib \assign tmp$ \; \DontPrintSemicolon
			$n \assign n - 1$ \;
		}
	} % end if
	\caption{Iterative Fibonacci program in our while language.}
	\label{alg:fib}
\end{algorithm}

\begin{figure}[!htb]
	\centering
	\input{./figures/tikz/cfg-fib.tex}
	\caption{Program graph of our iterative Fibonacci program (Algorithm~\ref{alg:fib}).}
	\label{fig:cfg-fib}
\end{figure}
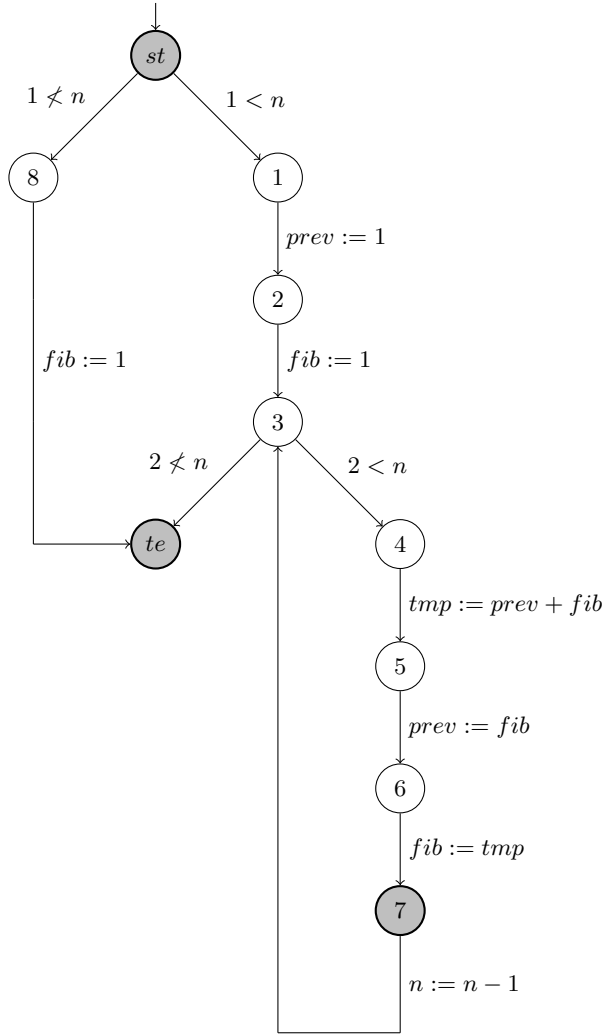

\begin{example}
	Algorithm~\ref{alg:fib} implements the computation of the $n$-th Fibonacci number for a given positive integer $n$ as a while program.
	This program serves as a running example throughout this paper.
	Figure~\ref{fig:cfg-fib} illustrates the program graph~\cite{Nielson2019} that corresponds to Algorithm~\ref{alg:fib}.
	Node $st$ represents its start and node $te$ its termination.
\end{example}

\section{Program Optimization by Aggregation}
\label{sec:decomp}
The key idea of our compilation approach is the radical decomposition of an input program into side-effect-free components that represent its control and data flow. 
This decomposition involves three main steps:
\begin{itemize}
\item path-wise separation of the conditional control flow and the computational aspect (Sec.~\ref{subsec:decomposition}), 
 \item decomposition of the program graph into acyclic fragments using cut points along the lines proposed in Floyd~\cite{floyd} (Sec.~\ref{subsec:cutpoints}), and
 \item path aggregation to cover an entire acyclic program fragment (Sec.~\ref{subsec:path_aggr}).
\end{itemize}
For every acyclic fragment, these first three steps result in an Algebraic Decision Diagram (ADD) whose terminal nodes determine the next effect on the program state and the point of continuation.
An Expression DAG that comprises all the computations required anywhere in the fragment eliminates redundant arithmetic expressions (Sec.~\ref{subsec:expr_dag}).
These two data structures together constitute the optimized program and allow for its rapid evaluation (Sec~\ref{subsec:exec}).

\subsection{Decomposition into Control and Data Flow}
\label{subsec:decomposition}
In the following, we show how to decompose a while program (Def.~\ref{def:while_language}) into its control and data flow. 
In order to achieve this and also reason about the correctness of our decomposition, we introduce a Structured Operational Semantics (SOS) that combines concrete and symbolic domains and which processes a tuple $\enangle{S, c, c_{_H} \sigma, \sigma_H}$ containing 
\begin{enumerate}
	\item a while program $S$ (Def.~\ref{def:while_language}),
	\item a Boolean variable $c$ called \emph{path-taken indicator} that states if the current path is the actually executed path w.r.t. the concrete program state (see 4.),
	\item a Boolean expression $c_{_H}$ (Def~\ref{def:expr}) called \emph{path condition} that symbolically aggregates branching conditions on the current path,
	\item a concrete program state $\sigma$ that stores the current (integer) values of program variables, and
	\item a symbolic program state $\sigma_{_H}$ called \emph{parallel assignment} that stores variable values as arithmetic expressions based on the initial program state.
\end{enumerate} 

The path-taken indicator $c$ (second component) and concrete program state $\sigma$ (fourth component) are only added to our representation in order to facilitate reasoning on the correctness of our approach.
Only the third and fifth components, namely the path condition and parallel assignment, are required for our decomposition-based compilation. 

We distinguish concrete and symbolic program states along with their corresponding semantics as follows:
\begin{definition}[Concrete vs. Symbolic State]
	Let $\var$ be a set of integer variables and $\ArEx$ denote arithmetic expressions (Def.~\ref{def:expr}) over symbolic versions of variables in $\var$. 
	Then a function $\sigma: \var \to \mathbb{Z}$ is called \emph{concrete program state} whereas a function $\sigma_{_H}: \var \to \ArEx$ is called \emph{symbolic program state}.
	
	For any expression $t \in (\ArEx \cup \BoEx)$, the syntax $\sem{t}(\sigma)$ denotes the evaluation of expression $t$ with respect to concrete program state $\sigma$, and $\symsem{t}(\sigma_{_H})$ the Herbrand interpretation~\cite{greibach1975theory,muller2005interprocedural} of expression $t$ with respect to symbolic state $\sigma_{_H}$.
	
	The semantic evaluation $\sem{\cdot}$ extends naturally to symbolic program states because the latter map variables to expressions. For any $x \in \var$, the resulting function is defined as follows:
	$$ \left( \sem{\sigma_{_H}}(\sigma)  \right)(x) \defined \sem{\sigma_{_H}(x)}(\sigma)   $$
	\label{def:concrete_vs_symbolic_state}
\end{definition}

\begin{example}
	Let $\var = \{ x,y  \}$  be a set of variables and $t = x + y$ a corresponding (arithmetic) expression. Given a concrete program state $\sigma = \{ x \mapsto 5, y \mapsto 4 \}$, we have $\sem{t}(\sigma) = 9$. Based on a symbolic program state $\sigma_{_H} = \{ x \mapsto 3 + 2, y \mapsto X \}$, we have $\symsem{t}(\sigma_{_H}) = (3 + 2) + X$. The latter is an expression over a symbolic version of variable $x \in \var$ and constants that is not evaluated further (Herbrand interpretation).
\end{example}

Note that the notions of symbolic execution~\cite{king1976symbolic} and iterated Herbrand interpretation are strongly related. 
We choose the latter for the presentation in this paper due to its clear formal roots. 
Other than symbolic execution, our main approach does not partially evaluate expressions, for example the sum of two integer constants. However, such optimizations can of course be incorporated into our compilation (see Sec.~\ref{subsec:normalization}).

Given our two domains of concrete and symbolic states, respectively, we now define our SOS with the following rules:
\begin{definition}[Combined-Domain SOS]
	For any (concrete or symbolic) program state $\sigma$, let $\sigma\{ y / x \}$ denote the substitution of $x$ by $y$ in $\sigma$. 
	Let a `~$\boldsymbol{\cdot}$' above a Boolean operator denote that this operation is evaluated semantically.
	Then our SOS is defined as follows:

  \normalfont
  \small
  \begin{equation*}
    \textrm{assign}~
    \frac{-}{
      \langle x \assign t, c, c_{_H}, \sigma, \sigma_{_H} \rangle
      ~ \longrightarrow ~
      \langle c, c_{_H}, \sigma \{ \sem{t}(\sigma) / x \}, \sigma_{_H} \{ \symsem{t}(\sigma_{_H}) / x \} \rangle} 
  \end{equation*}

  \begin{equation*}
    \textrm{skip}~
      \frac{-}{
      \langle \keyword{skip}, c, c_{_H}, \sigma, \sigma_{_H} \rangle
      ~ \longrightarrow ~
      \langle c, c_{_H}, \sigma, \sigma_{_H} \rangle}
	\end{equation*}
	
  \begin{equation*}
    \textrm{comp}_1~
    \frac{
      \langle S_1, c, c_{_H}, \sigma, \sigma_{_H} \rangle
      ~ \longrightarrow ~
      \langle S_1', c', c_{_H}', \sigma', \sigma_{_H}' \rangle}{
      \langle S_1; ~ S_2, c, c_{_H}, \sigma, \sigma_{_H} \rangle
      ~ \longrightarrow ~
      \langle S_1'; ~ S_2, c', c_{_H}',  \sigma', \sigma_{_H}' \rangle}
	\end{equation*}
	
	\begin{equation*}
    \textrm{comp}_2~
  	\frac{
      \langle S_1, c, c_{_H}, \sigma, \sigma_{_H} \rangle
      ~ \longrightarrow ~  
      \langle c',  c_{_H}', \sigma', \sigma_{_H}' \rangle}{
      \langle S_1;S_2, c, c_{_H}, \sigma, \sigma_{_H} \rangle
      ~ \longrightarrow ~  
      \langle S_2, c', c_{_H}',  \sigma', \sigma_{_H}' \rangle}
	\end{equation*}
	
	\begin{equation*}
    \textrm{if}_{\semTrue}~
    \frac{-}{
      \langle \keyword{if} ~ b ~ \{ S_1 \} ~ \keyword{else} ~ \{ S_2 \}, c, c_{_H}, \sigma, \sigma_{_H} \rangle
      ~ \longrightarrow ~
      \langle S_1, c \overset{\boldsymbol{\cdot}}{\land} \sem{b}(\sigma), c_{_H} \land \symsem{b}(\sigma_{_H}), \sigma, \sigma_{_H} \rangle}
	\end{equation*}
  
  \begin{equation*}
    \textrm{if}_{\semFalse}~
    \frac{-}{
      \langle \keyword{if} ~ b ~ \{ S_1 \} ~ \keyword{else} ~ \{ S_2 \}, c, c_{_H}, \sigma, \sigma_{_H} \rangle 
      ~ \longrightarrow ~  
      \langle S_2, c \overset{\boldsymbol{\cdot}}{\land} \overset{\boldsymbol{\cdot}}{\neg} \sem{b}(\sigma), c_{_H} \land \neg \symsem{b}(\sigma_{_H}), \sigma, \sigma_{_H} \rangle}
  \end{equation*}
  
  \begin{equation*}
    \textrm{wh}~
    \frac{-}{
      \langle \keyword{while} ~ b ~ \{ S \}, c, c_{_H}, \sigma, \sigma_{_H} \rangle
      ~ \longrightarrow ~
      \langle \keyword{if} ~ b ~ \{ S; ~ \keyword{while} ~ b ~ \{ S \} \} ~ \keyword{else} ~ \{ \keyword{skip} \}, c , c_{_H}, \sigma, \sigma_{_H} \rangle}
  \end{equation*}	
  
  \smallskip 
  \noindent As usual, the notation   
  $\enangle{S, c, c_{_H}, \sigma, \sigma_{_H}}
  \longrightarrow^* 
  \enangle{S', c', c_{_H}', \sigma', \sigma_{_H}'}
$
  denotes a sequence of SOS rule applications.
  
	\label{def:sos}
\end{definition}

\begin{example}
	Figure~\ref{fig:sym-exec-fib-n7} illustrates the application of our SOS rules based on the fragment
	\small{
    \begin{align*}
      &n \assign n - 1; ~ \keyword{while} ~ 2 < n ~ \{ tmp \assign prev + fib; ~ prev \assign fib; ~ fib \assign tmp; ~ n \assign n - 1 \}
    \end{align*}
	}
	of our Fibonacci program up to the program's termination (left branch) or a repetition of that fragment (right branch).
	Only the right branch would be taken based on the concrete program state.
	This fragment therefore represents the paths from node $7$ to nodes $te$ and $7$ in the program graph of Figure~\ref{fig:cfg-fib}.
	Corresponding program fragments (first component of the SOS tuple) are omitted in favor of edge annotations.
	Upper case letters represent initial symbolic values of program variables, e.g. $F$ is the initial symbolic value of $fib$.
	The difference between concrete and symbolic semantics becomes apparent when comparing the program states $\sigma$ and $\sigma_{_H}$.
	\label{example:sos}
\end{example}

\begin{figure}[!htb]
	\centering
  \input{./figures/tikz/sym-exec-fib-n7.tex}
	\caption{Execution of SOS rules (Def.~\ref{def:sos}) based on a loop program fragment of Algorithm~\ref{alg:fib}. This fragment extends from node $7$ to nodes $te$ (left branch) and $7$ (right branch) in Fig.~\ref{fig:cfg-fib}.}
	\label{fig:sym-exec-fib-n7}
\end{figure}

\begin{proposition}[Traditional SOS Incorporated]
	Our Combined-Domain SOS (Def.~\ref{def:sos}) is equivalent to the traditional `concrete' SOS~\cite{PLOTKIN} when
	\begin{enumerate}
		\item ignoring the components $c$, $c_{_H}$, and $\sigma_{_H}$ as well as
		\item adding side condition $\sem{b}(\sigma) \equiv \semTrue$ to rule $\textrm{if}_{\semTrue}$ and $\sem{b}(\sigma) \equiv \semFalse$ to rule $\textrm{if}_{\semFalse}$.
	\end{enumerate}
\label{proposition:traditional_sos_incorporated}
\end{proposition}
Note that a path would be part of the traditional `concrete' SOS iff $c = \semTrue$.
Based on these SOS rules, we can now state our main theorem about the correctness of our decomposition:
\begin{theorem}[Correctness of Control-Data-Decomposition]
  \label{thoerem:correctness_decomposition}
	Let $id$ denote the symbolic program state that maps each $v \in \var$ to its symbolic version.
	Let
  \begin{align*}
    \enangle{S, c, c_{_H}, \sigma, id}
    \longrightarrow^* 
    \enangle{S', c', c_{_H}', \sigma', \sigma_{_H}'}
  \end{align*}
	with $c = \sem{c_{_H}}(\sigma)$.
	Then the following holds for all expressions $t \in (\ArEx \cup \BoEx)$:
  \begin{align*}
    \sem{t}(\sigma') = \sem{\symsem{t}(\sigma_{_H}')}(\sigma)
  \end{align*}
\end{theorem}
Intuitively speaking, Theorem~\ref{thoerem:correctness_decomposition} states that the evaluation of an expression $t$ interpreted via $\sigma_{_H}$ with respect to the
initial state $\sigma$ is semantically equivalent to evaluating $t$ in the current state $\sigma'$.\footnote{This is an instance of a well-known substitution lemma.}
This theorem holds regardless of whether or not a given chain of SOS rule applications is part of the traditional `concrete' SOS semantics (Proposition~\ref{proposition:traditional_sos_incorporated}).
A corresponding proof follows from a straightforward induction over the SOS rules of Def.~\ref{def:sos} and the corresponding expressions (Def.~\ref{def:expr}).

The following corollary directly follows from Theorem~\ref{thoerem:correctness_decomposition} and the fact that our path-taken indicator $c$ could alternatively be stored as a Boolean variable in our program state: 

\begin{corollary}
  \label{corollary:correctness_decomposition}
	Let $\var$ be the set of program variables and let $id$ denote the symbolic program state that maps each $v \in \var$ to its symbolic version. 
	Let $\semTrue$ denote the semantic value of Boolean constant $true$ and let 
  \begin{align*}
    \enangle{S, \semTrue, true, \sigma, id}
    \longrightarrow^* 
    \enangle{S', c', c_{_H}', \sigma', \sigma_{_H}'} \textrm{.}
  \end{align*}
  Then, the following hold:\footnote{Please recall our definition of $\sem{ \sigma'_{_H}}(\sigma)$ from Def.~\ref{def:concrete_vs_symbolic_state}.} 
  \begin{enumerate}
    \item $c' = \sem{c'_{_H}}(\sigma)$
    \item $\sigma' = \sem{ \sigma'_{_H}}(\sigma)$
  \end{enumerate}
\end{corollary}

The first part of this corollary asserts that our path condition is correct. The second part is the reason why we call the symbolic program state \emph{parallel assignment}: Each variable can be updated independently of others because for each such $v \in \var$, the expression $\sigma_{_H}'(v)$ is only based on symbolic versions of variables in $\var$ and constants.
This side-effect-free update mechanism together with our aggregated path condition yields a decomposition of our program into data and control flow, respectively. 
In the following section, we will apply this decomposition to fragments of an input program.

\subsection{Segmentation into Acyclic Fragments}
\label{subsec:cutpoints}

In this section, we introduce how the decomposition of an input program into its control and data flow (Section~\ref{subsec:decomposition}) can be used for a compilation of that program.
By themselves, the SOS rules of Def.~\ref{def:sos} do not allow to statically transform a given program into a version where control and data are decomposed:
We do not have a bound for the length of an execution path and a static decomposition would therefore not terminate.

In order to circumvent this problem, we segment a given input program into acyclic fragments which we then decompose individually.
Our method is similar to the path variant of Floyd's inductive assertion method~\cite{floyd} for program verification: We choose a set of cut points such that they interrupt all loops in a given program. We choose as cut points
\begin{itemize}
 \item the start node,
 \item the termination node, and
 \item one node for each while construct
\end{itemize}
of a given while program (Def.~\ref{def:while_language}). 
Based on such a choice, every cut point
\begin{itemize}
	\item has a statically known number of outgoing paths to successor cut points, and
	\item these outgoing inter-cut point paths have a statically known length.
\end{itemize}
Because of these static bounds, a purely symbolic execution of such an acyclic fragment starting at a cut point $u$ is guaranteed to terminate. 
To execute a program fragment $S$ symbolically means to execute the SOS rules of Def.~\ref{def:sos} while (i) ignoring the path-taken indicator $c$ and concrete program state $\sigma$, and (ii) starting with $\enangle{S, \textvisiblespace, true, \textvisiblespace, id}$.

We can use these acyclic program fragments between cut points to execute the original program while preserving its semantics.
The following theorem asserts the correctness of our program segmentation:
\begin{theorem}[Compositionality]
	Let
	\begin{align*}
		\enangle{S,\semTrue, true, \sigma, id}
    &\longrightarrow^*
    \enangle{S', \semTrue, c_{_H}', \sigma', \sigma_{_H}'} , \\
		\enangle{S', \semTrue, true, \sigma', id}
    & \longrightarrow^*  
    \enangle{S'', \semTrue, c_{_H}'', \sigma'', \sigma_{_H}''}.
	\end{align*}
	Then the following holds:
  \begin{align*}
    \sigma'' = \sem{\sigma_{_H}''}(\sem{\sigma_{_H}'}(\sigma))
  \end{align*}
  \label{thoerem:correctness_segmentation}
\end{theorem}

This theorem follows straightforwardly from two applications of Corollary~\ref{corollary:correctness_decomposition}.

In a while program, every node in the corresponding program graph~\cite{Nielson2019} can be annotated with a fragment of that program\footnote{Note that due to the SOS rule for the while construct, a fragment of a program does not necessarily have to be a substring of that program.}. As a consequence, we can specify cut points visually as nodes in that graph.
\begin{example} 
	Figure~\ref{fig:cfg-fib} illustrates a choice of cut points for our Fibonacci program as nodes that are colored gray.
	In addition to start node $st$ and termination node $te$, we choose node $7$ as a cut point in order to interrupt the only while loop $3 \to 4 \to 5 \to 6 \to 7 \to 3$ in this program. The program fragment corresponding to cut point $7$ and the matching SOS rule application up to successor cut points were already introduced in Example~\ref{example:sos} and Figure~\ref{fig:sym-exec-fib-n7}.
\end{example}

The actual choice of cut points for a given input program is fundamental for later steps in our optimization process. 
This set of cut points therefore serves as a parameter of our optimization relative to which we can achieve optimal results, but which we choose heuristically.
Usually, the longer the fragments of a given program based on chosen cut points are, the more potential for optimization exists.

\begin{example}
	Note that in our Fibonacci program, choosing node $7$ (instead of $3$) to interrupt its loop results in the path $st \to 1 \to 2 \to 3 \to te$ to be kept uninterrupted by cut points. 
	This means that our approach allows to bypass a loop's cut point if that loop's body will not be entered.
\end{example}

\subsection{Path Aggregation Using Algebraic Decision Diagrams}
\label{subsec:path_aggr}
The result of our first two steps (Sec.~\ref{subsec:decomposition} and Sec.~\ref{subsec:cutpoints}) is a finite number of program fragments, each characterized by the symbolic execution traces up to their successor cut points: 

\begin{definition}[Contracted Cut Point Path]
  Let $u$ be a cut point associated with the program fragment $S$ and let $u'$ be one of its successor cut points associated with the program fragment $S'$. 
  Moreover, let 
  \begin{align*}
    \enangle{S, \textvisiblespace, true, \textvisiblespace, id}
    \longrightarrow^{*}
    \enangle{S', \textvisiblespace, c_{_H}, \textvisiblespace, \sigma_{_H}}.
  \end{align*}
  We call 
  \begin{align*}
    u \overset{c_{_H}, \sigma_{_H}}{\longrightarrow} u'
  \end{align*}
  a (contracted cut point) path from $u$ to $u'$ with the path condition $c_{_H}$ and the parallel assignment $\sigma_{_H}$.
\end{definition}

\begin{figure}[!htb]
  \centering
  \input{./figures/tikz/contr-fib.tex}
  \caption{Contracted cut point paths for our Fibonacci program (Algorithm~\ref{alg:fib}).}
  \label{fig:contr-fib}
\end{figure}
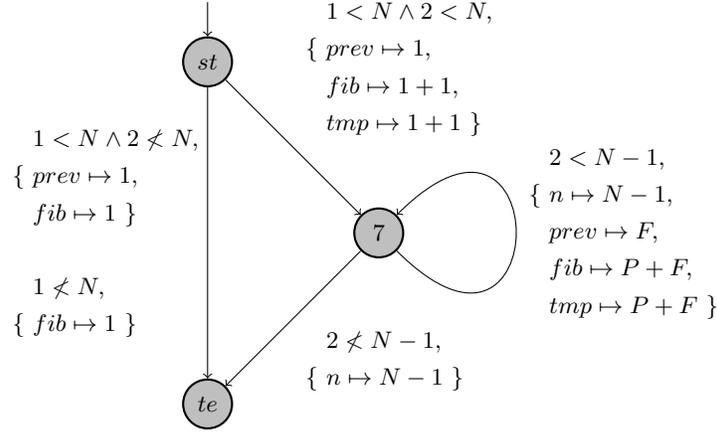
\begin{example}
  Figure~\ref{fig:contr-fib} illustrates this contracted view on our running example, the Fibonacci program (Algorithm~\ref{alg:fib}).
  For brevity, we omit all variables in the parallel assignments that remain unaffected, i.e. those that map variables to their respective symbolic version.
\end{example}

Because contracted cut point paths are free of side effects, the compositional aggregation of a cut point's outgoing paths yields a very natural fragment-wise transformation of the input program:
\begin{itemize}
  \item We will first transform these paths individually to Algebraic Decision Diagrams (ADDs).
  \item On this basis, we are able to aggregate them into a single ADD per cut point that completely defines the corresponding program fragment's behavior.
\end{itemize}

\subsubsection{Transformation of a Single Path}
Let us first consider a single path and its path condition.
This Boolean expression over atomic propositions (Def.~\ref{def:expr}) can be represented by means of a (Reduced Ordered) Binary Decision Diagram (BDD)~\cite{Bryant1986}.
For independent Boolean variables and a fixed variable order, this data structure
\begin{itemize}
  \item is a canonical representation for Boolean functions,
  \item realizes a minimal Binary Decision Diagram, and
  \item most importantly in this context, guarantees that every variable is encountered at most once when evaluating the diagram.
\end{itemize}

Because the atomic propositions that appear in our path conditions are not necessarily independent from one another, canonicity and minimality are only guaranteed on the level of their Herbrand interpretation.
Despite this, decision diagrams have proven to be an effective representation also for these more challenging problems.
Especially when combined with adequate treatment of infeasible paths~\cite{forest}, it is possible to tame their size and to benefit from their primary advantage: fast evaluation.

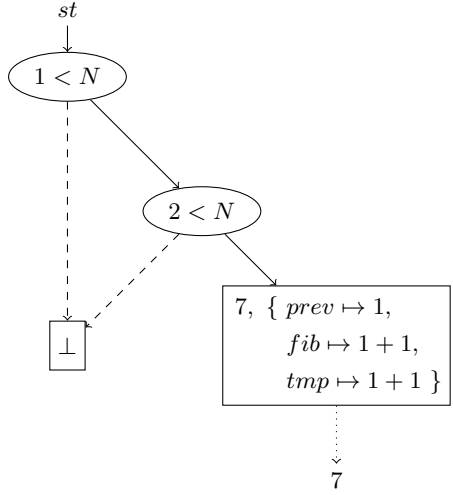
\begin{figure}
  \centering
  \input{./figures/tikz/add-fib-ns-n7.tex}
  \caption{Exemplary ADD for the contracted cut point path from node~$st$ to node~$7$ in Figure~\ref{fig:contr-fib} with the path condition $1 < N \wedge 2 < N$.}
  \label{fig:add-fib-st-n7}
\end{figure}
Figure~\ref{fig:add-fib-st-n7} visualizes the decision diagram that results from the exemplary path condition $1 < N \wedge 2 < N$ (cf. Fig.~\ref{fig:contr-fib}).
Note that this diagram is not exactly a BDD as its terminal nodes differ from Boolean values.
Instead, we use Algebraic Decision Diagrams (ADDs)~\cite{Bahar1997}, a generalization of this data structure, for two reasons:
\begin{itemize}
  \item For every path, we intend to store its effect on the program state, i.e. (i)~the parallel assignment and (ii)~the subsequent cut point, directly in the terminal nodes.
  \item We associate these decision diagrams with an algebraic structure that facilitates their composition.
  This is in fact key to our aggregation of a cut point's outgoing paths.
\end{itemize}

Both, parallel assignment and successor cut point, are applicable in case the path condition holds and they are irrelevant otherwise.
Hence, it is only natural to substitute them for the BDD's $1$-terminal.
For distinguishability, we also replace the BDD's $0$-terminal with a dedicated $\bot$ element.
The result is an ADD that is structurally analogous to the path condition BDD.

While BDDs can only represent Boolean functions of the form $\mathbb{B}^n \rightarrow \mathbb{B}$, ADDs allow for arbitrary co-domains.
At the same time, the desired properties of BDDs, i.e. canonicity, minimality, and fast evaluation, are fully preserved.

Moreover, we can lift algebraic operations that are defined on the co-domain to the level of their respective decision diagrams.
Similarly to the realization of standard Boolean operations on BDDs, efficient algorithms are known for ADDs with arbitrary algebraic structures.
Rather than conjunction, disjunction, and negation, we can, e.g., define a join operation on the co-domain.
The operation can then be applied to (i)~elements of the co-domain as well as to (ii)~the corresponding ADDs.
In this way, ADDs form an algebraic structure themselves, analogous to that of their co-domain.
For brevity, we denote the operations on the co-domain and those on the ADDs with the same symbols.

In our case, the algebraic structure is a lattice:

\begin{definition}[Path Aggregation Lattice]
  \label{def:lattice}
  Let $U$ be a set of cut points and let $\Sigma_{_H}$ be a set of all parallel assignments. 
  We define the path aggregation lattice $(A, \circ)$ on the carrier set 
  \begin{align*}
    A &:= \embrace{U \times \Sigma_{_H}} \cup \set{\bot, \top}
  \end{align*}
  with its supremum
  \begin{align*}
    a \circ b & :=
    \begin{cases}
      a & \textrm{if} ~ a = b \neq \bot \\
      a & \textrm{if} ~ b = \bot \neq a \\
      b & \textrm{if} ~ a = \bot \neq b \\
      \top & \textrm{otherwise.} \\
    \end{cases}
  \end{align*}
\end{definition}

$(A, \circ)$ forms a flat lattice with $\bot$ and $\top$ as its least and greatest element, respectively.
Note that $\top$ serves no purpose other than the natural completion of the structure and will never appear in our aggregation process.
The least element $\bot$ on the other hand does appear, but only in intermediate results.
Intuitively, we understand $\bot$ as the \emph{undefined} case in which the path condition does not hold.

The transformation of path conditions to BDDs and finally to ADDs constitutes a change in granularity.
Where predicates were previously expressed by means of possibly complex Boolean expressions ($\BoEx$ in Def.~\ref{def:expr}), they are now based on atomic propositions only ($\AP$ in Def.~\ref{def:expr}).
Any complexity of Boolean formulas beyond its $\AP$s is delegated to well-studied and efficient ADD algorithms in a service-oriented fashion.

\subsubsection{Aggregation of Multiple Paths}
To allow for a simultaneous evaluation of path conditions, we aggregate all outgoing paths per cut point.
With the supremum operation $\circ$ we can achieve this in a very simple way: We collect the most concrete information among these paths.
If either of them is undefined, i.e. $\bot$, we adopt the other, more concrete definition.

Being defined on their co-domain, we can easily apply $\circ$ to the previously constructed path ADDs:

\begin{definition}[Aggregated Path ADD]
  Let $u$ be a cut point, let $p_1, p_2, \ldots, p_n$ denote its outgoing paths, and let $dd$ denote the path transformation to equivalent ADDs.
  The aggregated path ADD for $u$ is the repeated application of $\circ$:
  \begin{align*}
    dd(u) &:= dd(p_1) \circ dd(p_2) \circ \cdots \circ dd(p_n)
  \end{align*}
\end{definition}

\begin{figure}
  \centering
  \subfloat[ADD for the start cut point $st$.]{
    \input{./figures/tikz/add-fib-ns.tex}
    \label{fig:add-fib-st}
  }
  \\
  \subfloat[ADD for the inner cut point $7$.]{
    \input{./figures/tikz/add-fib-n7.tex}
    \label{fig:add-fib-n7}
  }
  \caption{Aggregated path ADDs of the Fibonacci program.}
  \label{fig:add-fib}
\end{figure}
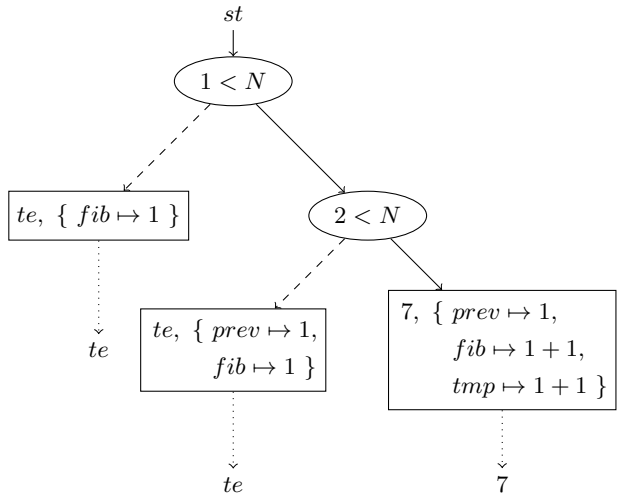
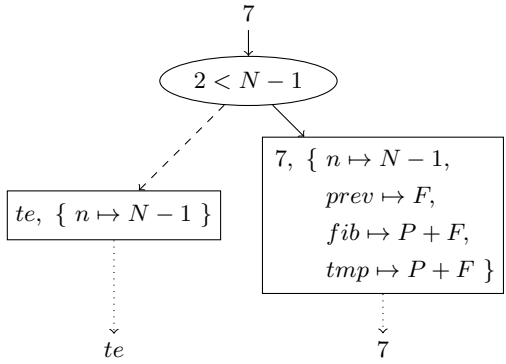
\begin{example}
  Figures~\ref{fig:add-fib-st} and~\ref{fig:add-fib-n7} show the final ADDs per cut point for our running example.
  Every non-termination cut point yields one ADD with exactly one terminal node for each of its successor cut points.
  Note that these ADDs are not necessarily trees but DAGs that share common sub-structures to keep the data structure small---an effect that becomes important for more complex path conditions.
\end{example}

Because the path conditions induce a partitioning on the Herbrand interpretation of atomic propositions, the $\top$ element will never appear in the aggregation process.
For the same reason, all undefined cases $\bot$ will eventually dissolve and the resulting ADDs yield only concrete pairs of parallel assignment and successor cut point.

With ADDs, we have found a program representation that allows to simultaneously evaluate all relevant path conditions.
At the same time, the program's semantics remain untouched.
Let $\sem{\cdot}_{_A}(\sigma)$ denote the standard semantic function to evaluate an ADD that, in our case, yields a parallel assignment based on some concrete program state $\sigma$ (Def.~\ref{def:concrete_vs_symbolic_state}).
The following theorem guarantees that we can represent a program fragment between cut points as an ADD:
\begin{theorem}[Correctness of Path Aggregation]
  \label{theo:path_aggr_correctness}
  Let $u$ be some cut point associated with the program fragment $S$, let $p_1, p_2, \ldots, p_n$ denote its outgoing paths, and let 
  \begin{align*}
    \enangle{S, \semTrue, true, \sigma, id}
    \longrightarrow^{*}
    \enangle{S', \semTrue, c_{_H}', \sigma', \sigma_{_H}'}.
  \end{align*}
  Then the following holds:\footnote{This theorem again uses our definition of $\sem{ \sigma'_{_H}}(\sigma)$ for a given parallel assignment (symbolic program state) $ \sigma'_{_H}$ (Def.~\ref{def:concrete_vs_symbolic_state}).} 
  \begin{align*}
   \sigma' = \sem{\sem{dd(p_1) \circ dd(p_2) \circ \cdots \circ dd(p_n)}_{_A}(\sigma)}(\sigma) \textrm{.}
  \end{align*}
\end{theorem}

Intuitively, Theorem~\ref{theo:path_aggr_correctness} justifies the use of the aggregated ADD representation to evaluate all relevant path conditions simultaneously.
Its proof is straightforward but tedious by induction over the aggregated paths and the ADD structure.

\subsection{Expression DAG}
\label{subsec:expr_dag}
The aggregation of paths into equivalent ADDs resolves redundancies among the path conditions per cut point. 
The problem of duplicate arithmetic expressions, however, remains. 
These appear not only in the parallel assignments of the ADDs' terminals but also in their inner nodes' predicates, i.e. in the atomic propositions. 
In fact, the duplication of arithmetic terms is a typical result of symbolic substitution, an operation that we heavily rely on during the course of our partial evaluation (Sec.~\ref{subsec:decomposition}). 
It is therefore crucial to also eliminate these redundancies. 

Achieving this goal is simple:
We resolve duplications in the form of an Expression DAG (ED).
Every constant and every program variable becomes a unique node---the atoms of this data structure. 
Based on these, the remaining expressions can be represented uniquely and with references to their respective sub-expressions. 

\begin{figure}
  \centering
  \input{./figures/tikz/ed-fib.tex}
  \caption{Expression DAG for our Fibonacci program (Algorithm~\ref{alg:fib}).}
  \label{fig:ed-fib}
\end{figure}
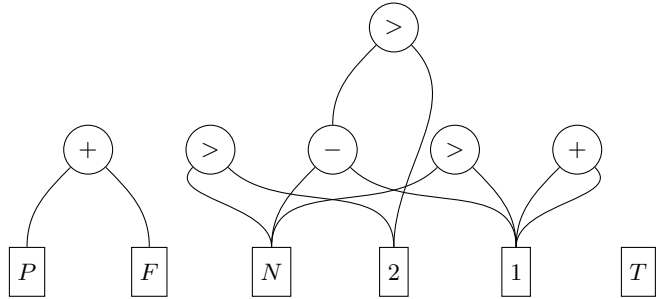
Figure~\ref{fig:ed-fib} shows the Expression DAG for our Fibonacci program (Algorithm~\ref{alg:fib}).
With additional optimizations (see Sec.~\ref{sec:opt}), especially loop unrolling and expression normalization, the number of expressions may grow quite drastically.
Using the ED is then crucial for the success of our program optimization.

\subsection{Execution of the Optimized Program}
\label{subsec:exec}
With Algebraic Decision Diagrams (Sec.~\ref{subsec:path_aggr}) and the Expression DAG (Sec.~\ref{subsec:expr_dag}), we have found an aggregated representation that finally reaches our goal: 
the efficient execution of a given program.

For any given cut point $u$ and concrete program state $\sigma$, we can execute the optimized representation of the corresponding program fragment:
We start at the root of $u$'s aggregated path ADD and evaluate its atomic propositions based on $\sigma$. 
The effect of the program fragment is not applied until a terminal is reached. 
Only then, the parallel assignment is applied to the program state $\sigma$. 
This execution is semantically equivalent to that of the original program fragment (Theorem~\ref{theo:path_aggr_correctness}). 

This fragment-wise execution is sufficient to also execute the original program in its entirety. 
Starting with some program state $\sigma$ at the initial cut point $st$, we effectively jump from cut point to cut point until the termination node $te$ is reached (see also Fig.~\ref{fig:contr-fib}). 
This iterative piecewise execution is then semantically equivalent to that of the entire input program (Theorem~\ref{thoerem:correctness_segmentation}).

\section{Additional Optimizations}
\label{sec:opt}
Our program decomposition into ADDs and an ED introduces a radically new compilation paradigm (Sec.~\ref{sec:decomp}).
It is not surprising that this method benefits from different optimization techniques than classical compilers do.
In this section, we present a variety of tailored optimizations and discuss their anticipated effect on our core approach.

\subsection{Loop Unrolling and Predicate Reordering}
\label{subsec:loop_unrolling}                                                                   
Our optimization is fundamentally based on the aggregation of acyclic program fragments and aims at processing large fragments (cf. Sec.~\ref{sec:intro}).
We can further increase the optimization capabilities of our decomposition by extending those fragments using loop unrolling, an optimization technique that we naturally support by relying on cut points.
One unrolled loop iteration in our representation yields one additional successor cut point in the corresponding ADD.

There exist different flavors of traditional loop unrolling, all of which involve an unroll factor $k$ of some sort. 
% A
On the one hand, while-loops can be unrolled by statically applying the while rule `wh' of Def.~\ref{def:sos} $k$ times~\cite{PLOTKIN,756775} (technique A). 
One benefit of this approach is a reduced number of jumps to the beginning of a while loop's body.
In addition, technique A maintains one homogeneous loop covering all step sizes up to the unrolling factor $k$.

% B
On the other hand, some approaches to the unrolling of for-loops copy $k$ iterations into a larger `big-step' loop~\cite{dongarra1979unrolling} (technique B). 
In the case where \emph{no} static loop bound is known, a separation of the number $n$ of iterations into $\lfloor \frac{n}{k} \rfloor$ big steps and the corresponding remainder is required.
The goal of such an `inner' unrolling is to speed up the given loop, e.g. by skipping a repetition of fine-grained control logic.

Our approach to loop unrolling is extremely generic:
Unrolling a loop once means to simply continue our decomposition that is based on a symbolic execution (Sec.~\ref{subsec:decomposition}) until we reach the same cut point again.
This loop unrolling combines accelerating effects of both techniques mentioned above while featuring additional benefits:
\begin{itemize}
	\item We reduce jumps to the loop body's entry point.
	
	\item Being based on a program graph, our cut-point-based approach can be applied to more general program languages involving not only while-loops, but e.g. also for-loops and goto statements.
	
	\item No explicit separation of the number of iterations into big steps and a remainder is required: Just like technique A, we unroll a loop for \emph{all} unroll factors between $0$ and a given maximum $k$ and store them in one homogeneous structure (an ADD).

	\item Our heuristics to flatten the ADDs imposes optimizations similar to binary search. In practice, we can randomize the predicate order of all non-loop-condition $\AP$s which (i)~usually results in rather balanced ADDs and (ii)~is completely generic, i.e. oblivious to specifics of loop unrolling.
\end{itemize}
Note that we never incur unnecessary overhead when evaluating $\AP$s that belong to different iterations of a loop condition based on unrolling: 
Regardless of their order in the ADD, sub-expressions are always shared in the ED (Sec.~\ref{subsec:expr_dag}). 
For example, the second and third unrolling of the loop condition in Algorithm~\ref{alg:fib} would read $2 < (n -1) - 1$ and $2 < ((n-1) -1) -1$, respectively.
If the right-hand side of the latter is evaluated first, then the right-hand side of the former---stored as a sub-DAG of the overall ED---is already evaluated.

Our evaluation in Section~\ref{sec:eval} illustrates the beneficial effects of loop unrolling and predicate reordering.

\subsection{Infeasible Path Elimination}
\label{subsec:unsat_path_removal}
% 
% Origin of infeasible paths. 
Similarly to the Random Forest compiler~\cite{forest,Breiman2001,Ho1995}, also here, infeasible path elimination has the potential to further simplify our aggregated program representation.
Loop unrolling and predicate reordering (Sec.~\ref{subsec:loop_unrolling}), in particular, yield infeasible paths and subsequent treatment can greatly reduce the diagrams' depths and sizes.  
In general, their origin is threefold: 
\begin{itemize}
  \item Already the input program may contain infeasible paths between its cut points, or even dead code. 
  \item As a result of loop unrolling, longer paths are considered, some of which may be infeasible. 
  \item Enforcing a particular predicate order swaps $\AP$s which, again, may result in infeasible paths---this time, however, only the aggregated path ADDs are affected. 
\end{itemize}

% Detect infeasible paths early.
In general, it is desirable to detect infeasibilities early in the optimization pipeline. 
Already at the stage of symbolic execution (Sec.~\ref{subsec:cutpoints}), we can validate path conditions for their satisfiability using an SMT solver. 
Rather than fully expanding the execution tree, we discard irrelevant paths early and effectively cut off all their continuations. 
This speeds up not only the optimization process, but, more importantly, yields smaller ADDs that are faster to evaluate. 
Those infeasible paths that occur only due to a change in the $\AP$ order appear first in the ADD. 
We eliminate these subsequently and directly in that data structure. 

% Merge paths. 
Because ADDs realize fully defined functions with regard to atomic propositions, we cannot remove infeasible paths from that data structure directly. 
Rather, we explicitly flag the outcome as undefined by replacing the corresponding terminal with $\bot$. 
Finally, the $\bot$ terminal can be dissolved similar to infeasible path elimination in~\cite{forest,ddds}. 
The sanitized ADD is a greatly simplified version of its original that differs only in cases that are infeasible anyways.

\subsection{Arithmetic Expression Normalization}
\label{subsec:normalization}
% 
% Simplify the many AEs.
The repeated symbolic substitution, especially through multiple iterations of a loop, generates a great amount of unique arithmetic expressions. 
Moving from a purely syntactic view to a more semantic understanding, we can exploit their inherent potential for simplification and condense these terms to a close to normal form. 
We delegate this generally non-trivial task to established SMT solving techniques. 

% Result.
As a result, the overall number of arithmetic operations is reduced quite drastically in many cases. 
This normalization also evaluates expressions already at compile time where possible. 

% Normalization method by example.
At its core, SMT technology essentially exploits common arithmetic laws\footnote{The exact normalization method should be chosen under consideration of a machine's number representation.} and, in case of the Fibonacci example, typically transforms expressions to their equivalent polynomials: 
\begin{align*}
  (((n - 1) - 1) - 1) &= n - 3 \\
  (P + F) + (F + (P + F)) &= 3F + 2P
\end{align*}

% Unify AEs for ADD and ED.
At the same time, this simplification serves a second purpose: 
Semantically equivalent expressions become unified, also syntactically.
This effect cascades to $\AP$s and reduces the overall number of syntactically different $\AP$s.
Finally, this translates to the corresponding ADDs where expression normalization allows for a smaller and shallower representation. 
Expression DAGs are affected in a similar fashion: 
Where the normalization unifies terms, it also merges nodes in the ED. 

\subsection{Impact on Our Fibonacci Example}
\begin{figure}
  \centering
  \subfloat[With $2$ loop unrollings, optimized $\AP$ order, and expression normalization.]{
    \input{./figures/tikz/add-fib-u2-n7.tex}
    \label{fig:add-fib-u2-n7}
  }
  \\
  \subfloat[With $2$ loop unrollings, optimized $\AP$ order, infeasible path elimination, and expression normalization.]{
    \input{./figures/tikz/add-fib-u2-sat-n7.tex}
    \label{fig:add-fib-u2-sat-n7}
  }
  \caption{Impact of additional optimizations on the aggregated path ADD.}
  \label{fig:add-fib-unrolled}
\end{figure}
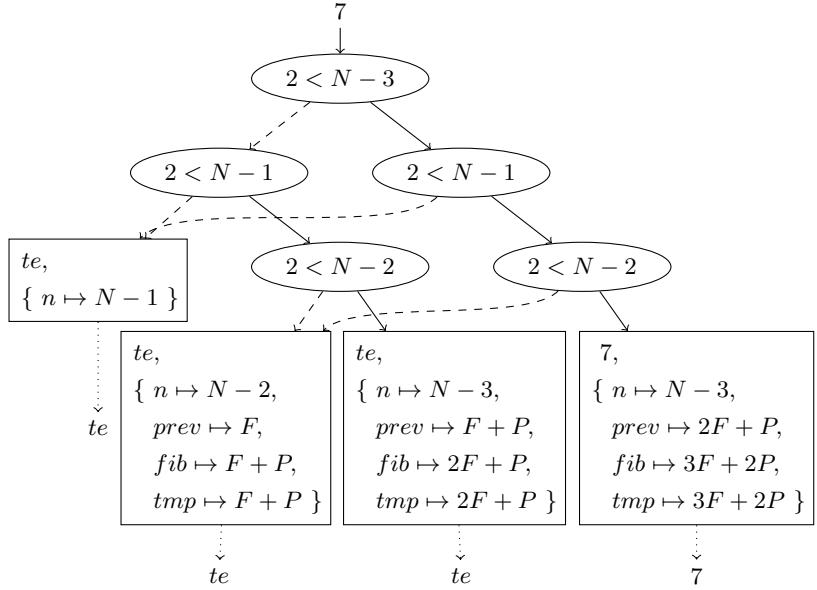
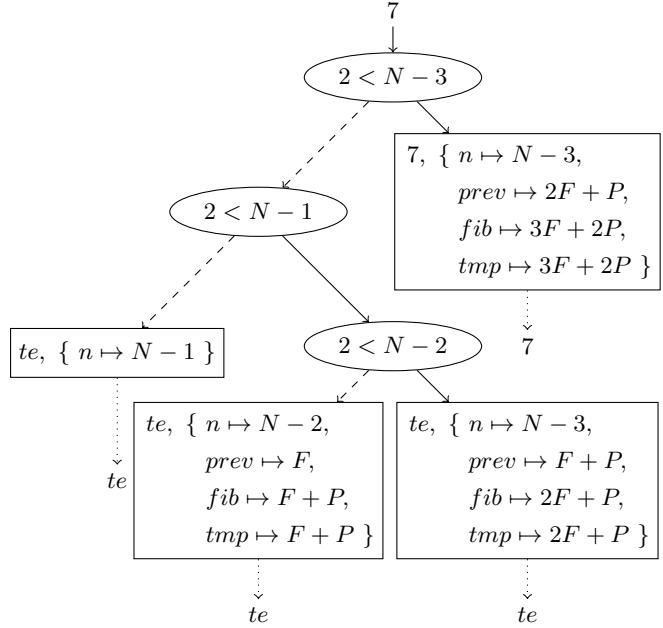
% 
% Fibonacci example. 
We have discussed four additional optimizations and their anticipated interplay with one another. 
Figure~\ref{fig:add-fib-u2-n7} shows the effect of loop unrolling, predicate reordering, and expression normalization on our Fibonacci program. 
In this example, the loop was unrolled twice and the $\AP$ order was chosen at random. 
Arithmetic expressions were condensed to simple forms such as $N - 3$ or $3F + 2P$, both of which were originally nested.

% Infeasible path elimination.
The resulting ADD contains two infeasible paths that, in this case, result solely from the loop unrolling. 
When $2 < N - 3$ holds, the truth value of the following two predicates $2 < N - 1$ and $2 < N - 2$ is already determined. 
Paths like these unnecessarily complicate the ADD and are, for this reason, subsequently eliminated. 
The result is shown in Figure~\ref{fig:add-fib-u2-sat-n7}: a smaller and shallower equivalent.
An important effect of the presented predicate order is that the biggest step through the unrolled loop can be evaluated with only one atomic proposition: $2 < N - 3$. 

\section{Evaluation}
\label{sec:eval}
We evaluate our aggregating program compilation (Sec.~\ref{sec:decomp}) along with additional optimizations (Sec.~\ref{sec:opt}) by examining execution times of our running example, a program that computes the $n$-Fibonacci number (Algorithm~\ref{alg:fib}).
Our goal for this evaluation is to be machine independent.
We therefore simulate execution time by counting costs of executed operations (Sec.~\ref{subsec:counting_costs}).
With all optimizations enabled, we are able to obtain a speedup of more than an order of magnitude compared to the original program when unrolling its loop up to $64$ times. Section~\ref{subsec:fib_eval} presents a detailed evaluation of the effects of different optimization techniques and lists key results of our evaluation.

\subsection{Measuring Execution Time by Counting Costs}
\label{subsec:counting_costs}
We measure the execution time of programs optimized using our approach in a machine-independent fashion based on the two data structures defined in Section \ref{sec:decomp}, i.e.\ our ADD and Expression DAG.
This independence is achieved by counting costs associated with executed operations. 
We associate a cost of $1$ with each
\begin{itemize}
	\item arithmetic operation,
	\item logical operation,
	\item comparison,
	\item conditional jump, and
	\item assignment.
\end{itemize} 
In the following, we briefly explain how costs were counted for the original input program and our optimized versions, respectively.

\noindent \textbf{Original Program:}
Our cost measurement is based on the program graph (Fig.~\ref{fig:cfg-fib}) of Algorithm~\ref{alg:fib} according to its natural semantics. 
The operations required to evaluate an expression ($\BoEx$ or $\ArEx$, see Def.~\ref{def:expr}) are based on a standard syntax tree.
Each branching node represents a conditional jump and therefore incurs an additional cost of $1$.

\noindent \textbf{Aggregated Program (ADD and ED):}
Our aggregated (optimized) program is executed as described in Section~\ref{subsec:exec}.
Here, we utilize our ED (Sec.~\ref{subsec:expr_dag}) and therefore evaluate the sub-DAG of this ED whose root is the evaluated expression. 
In addition, we count costs of $1$ for each visited inner node as it represents a conditional jump.
The costs for the parallel assignment located at a terminal of the ADD are counted as the sum of the costs of all its assignments. 

\subsection{Results}
\label{subsec:fib_eval}
Even though usual compilers hardly optimize our Fibonacci program (Algorithm~\ref{alg:fib}), the following evaluation shows that our aggregation-based compilation can drastically optimize this program and therefore significantly improve on state-of-the-art compilation techniques.

\begin{figure}[!htb]
	\centering
	\includegraphics[width=1\textwidth]{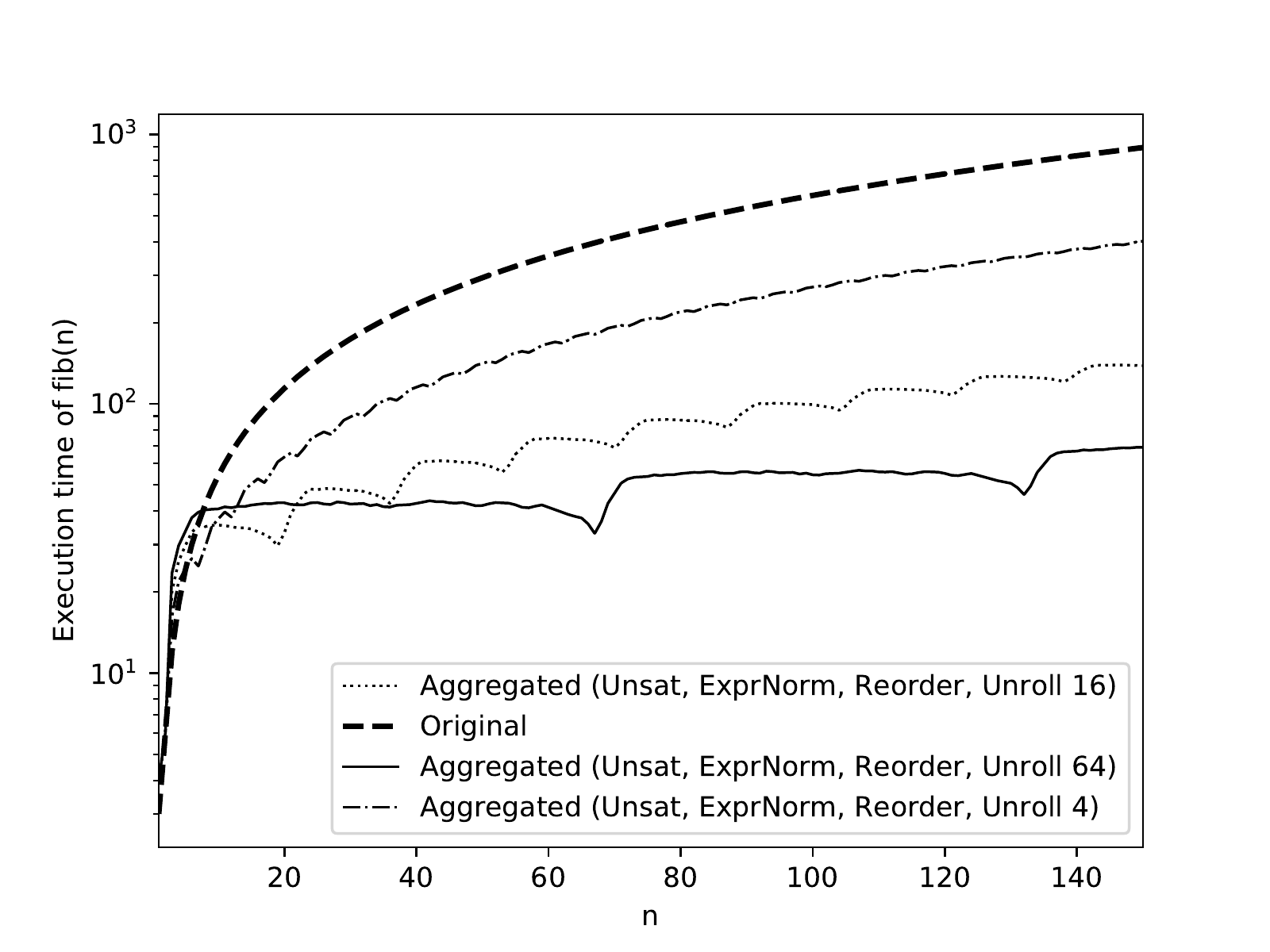}
	\caption{Execution times of versions with different numbers of loop unrolling while enabling all optimizations.}
	\label{fig:compare_lu}
\end{figure}

We first examine the execution time measured for the original program in comparison to optimized versions produced by our compiler in which the program's loop was unrolled up to $4,16,$ or $64$ times, respectively.
In cases in which our predicate reordering (Sec.~\ref{subsec:loop_unrolling}) that involves randomization is applied, we report the average of $1000$ unique measurements.
Figure \ref{fig:compare_lu} shows the execution times for the original program  and the aggregated versions that were optimized by our compiler using all techniques mentioned in Section~\ref{sec:opt}: expression normalization (Sec.~\ref{subsec:normalization}), infeasible path elimination (Sec.~\ref{subsec:unsat_path_removal}), and our predicate reordering (Sec.~\ref{subsec:loop_unrolling}) are applied.

Already in the case of $4$ loop unrollings, we can observe a declining execution time  in comparison to the original program.
Figure.~\ref{fig:compare_lu} clearly shows that variants in which a higher amount of loop unrolling was performed entail a larger speedup if $n$ exceeds the number of unrollings.
While for small $n$, loop unrolling might incur minor execution time overhead due to an increased average path length from the ADD's root to its terminals, this overhead is easily compensated by a drastic speedup for larger $n$.
For the case of $64$ loop unrollings and the computation of $fib(150)$, the measured execution time can be reduced by a factor of $13$ in comparison to the original program.

\begin{figure}[!htb]
	\centering
	\includegraphics[width=1\textwidth]{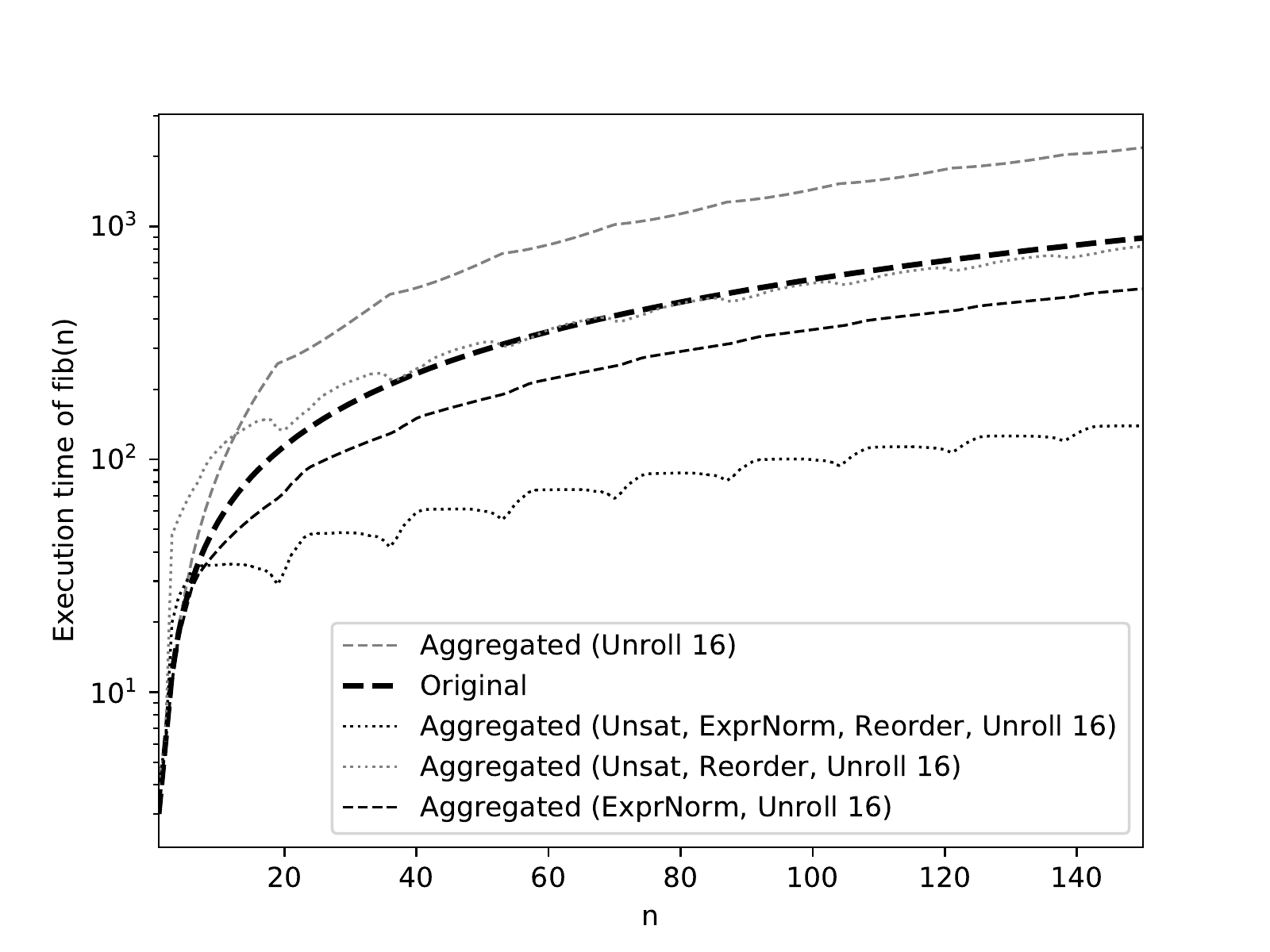}
	\caption{Execution times of different combinations of optimization techniques, all based on $16$ loop unrollings.}
	\label{fig:sequential_lu16}
\end{figure}

In contrast to the comparison in Figure \ref{fig:compare_lu} where all of our optimizations are enabled, Figure \ref{fig:sequential_lu16} shows how different combinations of optimizations affect execution time.
The fastest execution time is achieved when all optimizations are enabled.

Predicate reordering alone has a slightly negative effect as it introduces infeasible paths and might lead to a redundant evaluation of predicates.
Moreover, the infeasible path elimination alone has no effect as the original input program does not contain any infeasible paths.\footnote{For readability purposes, measurements during which only one of these optimizations is enabled are not represented in Figure~\ref{fig:sequential_lu16}.}
The combination of these two optimizations though is beneficial as the infeasible path elimination removes paths in the ADDs which are unnecessarily introduced by our predicate reordering.
Expression normalization turns out to be essential for the Fibonacci program as it simplifies the expanded expressions resulting from loop unrolling.
This is reflected in Figure \ref{fig:sequential_lu16} as the execution times of variants without expression normalization are higher compared to those where our expression normalization is enabled.

In summary, we have observed the following key results:
\begin{itemize}
	\item In the case of our Fibonacci program (Algorithm~\ref{alg:fib}), loop rolling enables the  full potential of additional optimizations and a large number of loop unrollings is preferable.
	\item Expression normalization accelerates the execution because it simplifies expanded expressions that result from loop unrolling.
	\item The combination of predicate reordering and infeasible path elimination is beneficial, whereas individually, these techniques are only helpful in specific cases, e.g.\ when the original input program already contains infeasible paths (cf.~\cite{forest}).
\end{itemize}

\section{Conclusion}
\label{sec:concl}

We have presented a new paradigm for program optimization which is based on aggressive aggregation, i.e. on a partial evaluation-based decomposition of acyclic program fragments into a pair of computationally optimal structures: 
an ADD to capture conditional branching and a parallel assignment that refers to an ED which realizes redundancy-free computation. 
This decomposition into side-effect-free representation of program fragments allows for powerful optimizations that semantically comprise effects traditionally aimed at by SSA form transformation, code specialization, common subexpression elimination, and (partial) redundancy elimination.

We have illustrated our approach along the optimization of the well-known iterative Fibonacci program, which, typically, is considered to lack any optimization potential, but could be `accelerated’ by more than an order of magnitude. 
The point here is that our technique supports loop unrolling as a first class optimization technique: 
It is tailored to optimally aggregate large program fragments---especially those resulting from multiple loop unrollings---while only incurring the expected linear increase in size. 

First experiments with parallel implementations indicate that our optimization technique is particularly well-suited to exploit parallelism, and that the simplicity of the resulting code eases the use of single instruction multiple data (SIMD) operations. 
We plan to further investigate this potential for parallelization.

Future work should combine our local aggregation-based optimization of program fragments with global program analyses for inter-program-fragment optimizations based on the contracted program graph (see Fig.~\ref{fig:contr-fib}).
Given the choice of cut points in our running example, one could e.g. observe that variable $tmp$ is never read in our aggregated program: Assignments to it are therefore dead code and can be removed.

Furthermore, we observed that our optimization serves as a drastic and redundancy-free obfuscation of the original code: 
Re-engineering the original program from our decomposed version seems almost impossible because both, the conditional structure in terms of an ADD---based on a mostly randomized predicate order---and the normalized ED, hardly give a hint about their origin.

\bibliographystyle{splncs04}
\bibliography{literature}
\end{document}

%% file: macros.tex
\newcommand{\set}[1]{\left \{ \; #1 \; \right \}}

\newcommand{\embrace}[1]{\left ( \; #1 \; \right )}
\newcommand{\enangle}[1]{\left \langle \; #1 \; \right \rangle}
\newcommand{\bnfdef}{~~ ::= ~~}
\newcommand{\bnfor}{~~ | ~~}

\newcommand{\sem}[1]{\llbracket \; {#1} \; \rrbracket}
\newcommand{\symsem}[1]{\sem{#1}_{_H}}
\newcommand{\semTrue}{\mathord{\mathit{tt}}}
\newcommand{\semFalse}{\mathord{\mathit{ff}}}
\newcommand{\defined}{=_{\textit{def}}~}

\newcommand{\var}{\text{\upshape V}}
\newcommand{\assign}{:=}
\newcommand{\keyword}[1]{\textbf{#1}}
\newcommand{\ArEx}{\text{\upshape AE}}
\newcommand{\BoEx}{\text{\upshape BE}}
\newcommand{\AP}{\text{\upshape AP}}

%% file: figures/tikz/tikz.tex
\usetikzlibrary{shapes, arrows}
\tikzstyle{cfg} = [x=5em, y=5em]
\tikzstyle{cfg-arrow} = [->]
\tikzstyle{cfg-arrow-pre} = []
\tikzstyle{cfg-node} = [circle, draw, minimum height=2em, text centered]
\tikzstyle{cfg-node-emph} = [circle, draw, thick, fill=lightgray, minimum height=2em, text centered]
\tikzstyle{sym-exec} = [x=7em, y=7em]
\tikzstyle{sym-exec-arrow} = [->]
\tikzstyle{sym-exec-node} = [rectangle, draw, minimum height=2em, text centered]
\tikzstyle{dd} = [x=5.5em, y=5.5em]
\tikzstyle{dd-arrow} = [->]
\tikzstyle{dd-arrow-then} = [->]
\tikzstyle{dd-arrow-else} = [dashed, ->]
\tikzstyle{dd-arrow-continuation} = [dotted, ->]
\tikzstyle{dd-node} = []
\tikzstyle{dd-node-int} = [ellipse, draw, minimum height=2em, text centered]
\tikzstyle{dd-node-term} = [rectangle, draw, minimum height=2em, text centered]
\tikzstyle{ed} = [x=5em, y=5em]
\tikzstyle{ed-arrow} = []
\tikzstyle{ed-node-int} = [circle, draw, minimum width=2em, text centered]
\tikzstyle{ed-node-term} = [rectangle, draw, minimum height=2em, text centered]

%% file: figures/tikz/cfg-fib.tex
\begin{tikzpicture}[cfg]
  \node (init) [] at (0,0.5) {};
  \node (st) [cfg-node-emph] at (0,0) {$st$};
  \node (n1) [cfg-node] at (1,-1) {$1$};
  \node (n2) [cfg-node] at (1,-2) {$2$};
  \node (n3) [cfg-node] at (1,-3) {$3$};
  \node (n4) [cfg-node] at (2,-4) {$4$};
  \node (n5) [cfg-node] at (2,-5) {$5$};
  \node (n6) [cfg-node] at (2,-6) {$6$};
  \node (n7) [cfg-node-emph] at (2,-7) {$7$};
  \node (te) [cfg-node-emph] at (0,-4) {$te$};
  \node (n8) [cfg-node] at (-1,-1) {$8$};
  \draw [cfg-arrow] (init) to node {} (st);
  \draw [cfg-arrow] (st) -- node [anchor=south west] {$1 < n$} (n1);
  \draw [cfg-arrow] (n1) -- node [anchor=west] {$prev := 1$} (n2);
  \draw [cfg-arrow] (n2) -- node [anchor=west] {$fib := 1$} (n3);
  \draw [cfg-arrow] (n3) -- node [anchor=south west] {$2 < n$} (n4);
  \draw [cfg-arrow] (n4) -- node [anchor=west] {$tmp := prev + fib$} (n5);
  \draw [cfg-arrow] (n5) -- node [anchor=west] {$prev := fib$} (n6);
  \draw [cfg-arrow] (n6) -- node [anchor=west] {$fib := tmp$} (n7);
  \draw [cfg-arrow-pre] (n7) -- node [anchor=west] {$n := n - 1$} (2,-8);
  \draw [cfg-arrow-pre] (2,-8) -- (1,-8);
  \draw [cfg-arrow] (1,-8) -- (n3);
  \draw [cfg-arrow] (n3) -- node [anchor=south east] {$2 \nless n$} (te);
  \draw [cfg-arrow] (st) -- node [anchor=south east] {$1 \nless n$} (n8);
  \draw [cfg-arrow-pre] (n8) -- (-1,-2);
  \draw [cfg-arrow-pre] (-1,-2) -- node [anchor=west] {$fib := 1$} (-1,-3);
  \draw [cfg-arrow-pre] (-1,-3) -- (-1,-4);
  \draw [cfg-arrow] (-1,-4) -- (te);
\end{tikzpicture}

%% file: figures/tikz/sym-exec-fib-n7.tex
\begin{tikzpicture}[sym-exec, x=9em, y=9em]
  \node (n7) [sym-exec-node] at (0,0) {$\begin{aligned}c &= \semTrue, \\ %
  c_{H} &= true, \\ %
  \sigma &= \{ ~ n \mapsto 4, prev \mapsto 5, fib \mapsto 8, tmp \mapsto 8 ~ \} \\ % 
  \sigma_{H} &= \{ ~ n \mapsto N, prev \mapsto P, fib \mapsto F, tmp \mapsto T ~ \}\end{aligned}$};
  \node (n3) [sym-exec-node] at (0,-1) {$\begin{aligned}c &= \semTrue, \\ %
  c_{H} &= true, \\ %
  \sigma &= \{ ~ n \mapsto 3, \ldots ~ \} \\ % 
  \sigma_H &= \{ ~ n \mapsto N - 1, \ldots ~ \}\end{aligned}$};
  \node (n4) [sym-exec-node] at (1,-2) {$\begin{aligned}c &= \semTrue, \\ %
  c_{H} &= 2 < N - 1, \\ %
  \sigma &= \{ ~ \ldots, prev \mapsto 5, fib \mapsto 8, \ldots ~ \} \\ % 
  \sigma_H &= \{ ~ \ldots, prev \mapsto P, fib \mapsto F, \ldots ~ \}\end{aligned}$};
  \node (n5) [sym-exec-node] at (1,-3) {$\begin{aligned}c &= \semTrue, \\ %
  c_{H} &= 2 < N - 1, \\ %
  \sigma &= \{ ~ \ldots, fib \mapsto 8, tmp \mapsto 13 ~ \} \\ % 
  \sigma_H &= \{ ~ \ldots, fib \mapsto F, tmp \mapsto P + F ~ \}\end{aligned}$};
  \node (n6) [sym-exec-node] at (1,-4) {$\begin{aligned}c &= 1, \\ %
  c_{H} &= 2 < N - 1, \\ %
  \sigma &= \{ ~ \ldots, prev \mapsto 8, tmp \mapsto 13 ~ \} \\ % 
  \sigma_H &= \{ ~ \ldots, prev \mapsto F, tmp \mapsto P + F ~ \}\end{aligned}$};
  \node (n7') [sym-exec-node] at (1,-5) {$\begin{aligned}c &= 1, \\ %
  c_{H} &= 2 < N - 1, \\ %
  \sigma &= \{ ~ \ldots, fib \mapsto 13, \ldots ~ \} \\ % 
  \sigma_H &= \{ ~ \ldots, fib \mapsto P + F, \ldots ~ \}\end{aligned}$};
  % \node (d1) at (1,-6) {$\vdots$};
  \node (nt) [sym-exec-node] at (-1,-2) {$\begin{aligned}c &= \semFalse, \\ %
  c_{H} &= 2 \nless N - 1, \\ %
  \sigma &= \{ ~ \ldots ~ \} \\ % 
  \sigma_H &= \{ ~ \ldots ~ \}\end{aligned}$};
  \draw [sym-exec-arrow] (n7) -- node [anchor=west] {$n := n - 1$} (n3);
  \draw [sym-exec-arrow] (n3) -- node [anchor=west, xshift=.5em] {$2 < n$} (n4);
  \draw [sym-exec-arrow] (n4) -- node [anchor=west] {$tmp := prev + fib$} (n5);
  \draw [sym-exec-arrow] (n5) -- node [anchor=west] {$prev := fib$} (n6);
  \draw [sym-exec-arrow] (n6) -- node [anchor=west] {$fib := tmp$} (n7');
  % \draw [sym-exec-arrow] (n7') -- (d1);
  \draw [sym-exec-arrow] (n3) -- node [anchor=east, xshift=-.5em] {$2 \nless n$} (nt);
\end{tikzpicture}

%% file: figures/tikz/contr-fib.tex
\begin{tikzpicture}[cfg, x=7em, y=7em]
  \node (init) [] at (-1,1.4) {};
  \node (ns) [cfg-node-emph] at (-1,1) {$st$};
  \node (n7) [cfg-node-emph] at (0,0) {$7$};
  \node (nt) [cfg-node-emph] at (-1,-1) {$te$};
  \draw [cfg-arrow] (init) to node {} (ns);
  \draw [cfg-arrow] (ns) to node [anchor=south west] {$\begin{aligned}&1 < N \wedge 2 < N, \\ \{ ~ &prev \mapsto 1, \\ &fib \mapsto 1 + 1, \\ &tmp \mapsto 1 + 1 ~ \}\end{aligned}$} (n7);
  \draw [cfg-arrow] (ns) to node [anchor=east] {$\begin{aligned}&1 < N \wedge 2 \nless N, \\ \{ ~ &prev \mapsto 1, \\ &fib \mapsto 1 ~ \} \\ \\ &1 \nless N, \\ \{ ~ &fib \mapsto 1 ~ \}\end{aligned}$} (nt);
  \draw [cfg-arrow] (n7) to [in=45, out=-45, looseness=16] node [anchor=west] {$\begin{aligned}&2 < N - 1, \\ \{ ~ &n \mapsto N - 1, \\ &prev \mapsto F, \\ &fib \mapsto P + F, \\ &tmp \mapsto P + F ~ \}\end{aligned}$} (n7);
  \draw [cfg-arrow] (n7) to node [anchor=north west] {$\begin{aligned}&2 \nless N - 1, \\ \{ ~ &n \mapsto N - 1 ~ \}\end{aligned}$} (nt);
\end{tikzpicture}

%% file: figures/tikz/add-fib-ns-n7.tex
\begin{tikzpicture}[dd]
  \node (st) [dd-node] at (0,-0.5) {$st$};
  \node (i1) [dd-node-int] at (0,-1) {$1 < N$};
  \node (i2) [dd-node-int] at (1,-2) {$2 < N$};
  \node (t2) [dd-node-term] at (2,-3) {$\begin{aligned}7, ~ \{ ~ &prev \mapsto 1, \\ &fib \mapsto 1 + 1, \\ &tmp \mapsto 1 + 1 ~ \}\end{aligned}$};
  \node (n7') [dd-node] at (2,-4) {$7$};
  \node (t1) [dd-node-term] at (0,-3) {$\bot$};
  \draw [dd-arrow] (st) -- (i1);
  \draw [dd-arrow-then] (i1) -- (i2);
  \draw [dd-arrow-then] (i2) -- (t2);
  \draw [dd-arrow-else] (i2) -- (t1);
  \draw [dd-arrow-else] (i1) -- (t1);
    \draw [dd-arrow-continuation] (t2) -- (n7');
\end{tikzpicture}

%% file: figures/tikz/add-fib-ns.tex
\begin{tikzpicture}[dd]
  \node (ns) [dd-node] at (0,-0.5) {$st$};
  \node (i1) [dd-node-int] at (0,-1) {$1 < N$};
  \node (i2) [dd-node-int] at (1,-2) {$2 < N$};
  \node (t3) [dd-node-term] at (2,-3) {$\begin{aligned}7, ~ \{ ~ &prev \mapsto 1, \\ &fib \mapsto 1 + 1, \\ &tmp \mapsto 1 + 1 ~ \}\end{aligned}$};
  \node (n7) [dd-node] at (2,-4) {$7$};
  \node (t2) [dd-node-term] at (0,-3) {$\begin{aligned}te, ~ \{ ~ &prev \mapsto 1, \\ &fib \mapsto 1 ~ \}\end{aligned}$};
  \node (nt) [dd-node] at (0,-4) {$te$};
  \node (t1) [dd-node-term] at (-1,-2) {$te, ~ \{ ~ fib \mapsto 1 ~ \}$};
  \node (nt') [dd-node] at (-1,-3) {$te$};
  \draw [dd-arrow] (ns) -- (i1);
  \draw [dd-arrow-then] (i1) -- (i2);
  \draw [dd-arrow-then] (i2) -- (t3);
  \draw [dd-arrow-continuation] (t3) -- (n7);
  \draw [dd-arrow-else] (i2) -- (t2);
  \draw [dd-arrow-continuation] (t2) -- (nt);
  \draw [dd-arrow-else] (i1) -- (t1);
  \draw [dd-arrow-continuation] (t1) -- (nt');
\end{tikzpicture}

%% file: figures/tikz/add-fib-n7.tex
\begin{tikzpicture}[dd]
  \node (n7) [dd-node] at (0,-0.5) {$7$};
  \node (i1) [dd-node-int] at (0,-1) {$2 < N - 1$};
  \node (t2) [dd-node-term] at (1,-2) {$\begin{aligned}7, ~ \{ ~ &n \mapsto N - 1, \\ %
  &prev \mapsto F, \\ %
  &fib \mapsto P + F, \\ %
  &tmp \mapsto P + F ~ \}\end{aligned}$};
  \node (n7') [dd-node] at (1,-3) {$7$};
  \node (t1) [dd-node-term] at (-1,-2) {$te, ~ \{ ~ n \mapsto N - 1 ~ \}$};
  \node (nt) [dd-node] at (-1,-3) {$te$};
  \draw [dd-arrow] (n7) -- (i1);
  \draw [dd-arrow-then] (i1) -- (t2);
  \draw [dd-arrow-continuation] (t2) -- (n7');
  \draw [dd-arrow-else] (i1) -- (t1);
  \draw [dd-arrow-continuation] (t1) -- (nt);
\end{tikzpicture}

%% file: figures/tikz/ed-fib.tex
\begin{tikzpicture}[ed]
  \node (prev) [ed-node-term] at (0,0) {$P$};
  \node (fib) [ed-node-term] at (1,0) {$F$};
  \node (n) [ed-node-term] at (2,0) {$N$};
  \node (c2) [ed-node-term] at (3,0) {$2$};
  \node (c1) [ed-node-term] at (4,0) {$1$};
  \node (tmp) [ed-node-term] at (5,0) {$T$};
  \node (i5) [ed-node-int] at (0.5,1) {$+$};
  \node (i1) [ed-node-int] at (1.5,1) {$>$};
  \node (i2) [ed-node-int] at (2.5,1) {$-$};
  \node (i3) [ed-node-int] at (3.5,1) {$>$};
  \node (i4) [ed-node-int] at (4.5,1) {$+$};
  \node (i6) [ed-node-int] at (3,2) {$>$};
  \draw [ed-arrow] (n) to [out=90, in=225] (i1);
  \draw [ed-arrow] (c2) to [out=90, in=315] (i1);
  \draw [ed-arrow] (n) to [out=90, in=225] (i2);
  \draw [ed-arrow] (c1) to [out=90, in=315] (i2);
  \draw [ed-arrow] (n) to [out=90, in=225] (i3);
  \draw [ed-arrow] (c1) to [out=90, in=315] (i3);
  \draw [ed-arrow] (c1) to [out=90, in=225] (i4);
  \draw [ed-arrow] (c1) to [out=90, in=315] (i4);
  \draw [ed-arrow] (prev) to [out=90, in=225] (i5);
  \draw [ed-arrow] (fib) to [out=90, in=315] (i5);
  \draw [ed-arrow] (i2) to [out=90, in=225] (i6);
  \draw [ed-arrow] (c2) to [out=90, in=315] (i6);
\end{tikzpicture}

%% file: figures/tikz/add-fib-u2-n7.tex
\begin{tikzpicture}[dd]
  \node (n7) [dd-node] at (0,-0.2) {$7$};
  \node (i1) [dd-node-int] at (0,-.7) {$2 < N - 3$};
  \node (i2) [dd-node-int] at (0.9,-1.4) {$2 < N - 1$};
  \node (i3) [dd-node-int] at (1.8,-2.1) {$2 < N - 2$};
  \node (t1) [dd-node-term] at (2.65,-3.3) {$\begin{aligned}7&, \\ 
  	\{ ~ &n \mapsto N - 3, \\ %
  &prev \mapsto 2 F + P, \\ %
  &fib \mapsto 3 F + 2 P, \\ %
  &tmp \mapsto 3 F + 2 P ~ \}\end{aligned}$};
  \node (n7') [dd-node] at (2.65,-4.4) {$7$};
  \node (t2) [dd-node-term] at (-0.85,-3.3) {$\begin{aligned} te&, \\ 
  	\{ ~ &n \mapsto N - 2, \\ %
  &prev \mapsto F, \\ %
  &fib \mapsto F + P, \\ %
  &tmp \mapsto F + P ~ \} \end{aligned}$};
  \node (nt) [dd-node] at (-0.9,-4.4) {$te$};
  \node (t3) [dd-node-term] at (-1.8,-2.2) {$\begin{aligned}
  	te&, \\
  	\{ ~ &n \mapsto N - 1 ~ \}
  	\end{aligned}$};
  \node (nt') [dd-node] at (-1.8,-3.3) {$te$};
  \node (t4) [dd-node-term] at (0.85,-3.3) {$ \begin{aligned} te&, \\ 
  	\{ ~ &n \mapsto N - 3, \\ %
  &prev \mapsto F + P, \\ %
  &fib \mapsto 2 F + P, \\ %
  &tmp \mapsto 2 F + P ~ \} \end{aligned} $};
  \node (nt'') [dd-node] at (0.9,-4.4) {$te$};
  \node (i4) [dd-node-int] at (-0.9,-1.4) {$2 < N - 1$};
  \node (i5) [dd-node-int] at (0,-2.1) {$2 < N - 2$};
  \draw [dd-arrow] (n7) -- (i1);
  \draw [dd-arrow-then] (i1) -- (i2);
  \draw [dd-arrow-then] (i2) -- (i3);
  \draw [dd-arrow-then] (i3) -- (t1);
  \draw [dd-arrow-continuation] (t1) -- (n7');
  \draw [dd-arrow-else] (i3) to [in=45, out=225, looseness=0.5] (t2);
  \draw [dd-arrow-continuation] (t2) -- (nt);
  \draw [dd-arrow-else] (i2) to [in=45, out=225, looseness=0.5] (t3);
  \draw [dd-arrow-continuation] (t3) -- (nt');
  \draw [dd-arrow-else] (i1) -- (i4);
  \draw [dd-arrow-then] (i4) -- (i5);
  \draw [dd-arrow-then] (i5) -- (t4);
  \draw [dd-arrow-continuation] (t4) -- (nt'');
  \draw [dd-arrow-else] (i5) -- (t2);
  \draw [dd-arrow-else] (i4) -- (t3);
\end{tikzpicture}

%% file: figures/tikz/add-fib-u2-sat-n7.tex
\begin{tikzpicture}[dd]
  \node (n7) [dd-node] at (0,-0.5) {$7$};
  \node (i1) [dd-node-int] at (0,-1) {$2 < N - 3$};
  \node (t1) [dd-node-term] at (1,-2) {$\begin{aligned}7, ~ \{ ~ &n \mapsto N - 3, \\ %
  &prev \mapsto 2 F + P, \\ %
  &fib \mapsto 3 F + 2 P, \\ %
  &tmp \mapsto 3 F + 2 P ~ \}\end{aligned}$};
  \node (n7') [dd-node] at (1,-3) {$7$};
  \node (t2) [dd-node-term] at (-1,-4) {$\begin{aligned} te, ~ \{ ~ &n \mapsto N - 2, \\ %
  &prev \mapsto F, \\ %
  &fib \mapsto F + P, \\ %
  &tmp \mapsto F + P ~ \} \end{aligned}$};
  \node (nt) [dd-node] at (-1,-5) {$te$};
  \node (t3) [dd-node-term] at (-2.05,-3.05) {$te, ~ \{ ~ n \mapsto N - 1 ~ \}$};
  \node (nt') [dd-node] at (-2.05,-4) {$te$};
  \node (t4) [dd-node-term] at (1,-4) {$ \begin{aligned} te, ~ \{ ~ &n \mapsto N - 3, \\ %
  &prev \mapsto F + P, \\ %
  &fib \mapsto 2 F + P, \\ %
  &tmp \mapsto 2 F + P ~ \} \end{aligned} $};
  \node (nt'') [dd-node] at (1,-5) {$te$};
  \node (i4) [dd-node-int] at (-1,-2) {$2 < N - 1$};
  \node (i5) [dd-node-int] at (0,-3) {$2 < N - 2$};
  \draw [dd-arrow] (n7) -- (i1);
  \draw [dd-arrow-then] (i1) -- (t1);
  \draw [dd-arrow-continuation] (t1) -- (n7');
  \draw [dd-arrow-continuation] (t2) -- (nt);
  \draw [dd-arrow-continuation] (t3) -- (nt');
  \draw [dd-arrow-else] (i1) -- (i4);
  \draw [dd-arrow-then] (i4) -- (i5);
  \draw [dd-arrow-then] (i5) -- (t4);
  \draw [dd-arrow-continuation] (t4) -- (nt'');
  \draw [dd-arrow-else] (i5) -- (t2);
  \draw [dd-arrow-else] (i4) -- (t3);
\end{tikzpicture}